\documentclass[twocolumn]{aa}
\usepackage{graphics}
\usepackage{graphicx}
\usepackage{amsfonts}
\include{epsf}
\include{epstopdf}
\DeclareGraphicsExtensions{.eps,.jpg,.pdf}
\newcommand{\bfo}[1]{\mbox{\boldmath $#1$}}
\def\bvarphi{\mbox{\boldmath $\varphi$}}

\begin{document}
\newcommand{\beq}{\begin{equation}}
\newcommand{\eeq}{\end{equation}}
%****** Begin Definitions *************************
%\def {\sffamily}
\def\la{\hbox{\raise.35ex\rlap{$<$}\lower.6ex\hbox{$\sim$}\ }}
\def\ga{\hbox{\raise.35ex\rlap{$>$}\lower.6ex\hbox{$\sim$}\ }}
\def\runit{\hat {\bf  r}}
\def\phunit{\hat {\bfo \bvarphi}}
\def\etaunit{\hat {\bfo \eta}}
\def\zunit{\hat {\bf z}}
\def\zetaunit{\hat {\bfo \zeta}}
\def\xiunit{\hat {\bfo \xi}}
\def\beq{\begin{equation}}
\def\eeq{\end{equation}}
\def\beqa{\begin{eqnarray}}
\def\eeqa{\end{eqnarray}}
\def\sub#1{_{_{#1}}}
\def\order#1{{\cal O}\left({#1}\right)}
\newcommand{\sfrac}[2]{\small \mbox{$\frac{#1}{#2}$}}
%****** End Definitions ***************************
%
%
\title{{Potential vorticity dynamics in the framework of disk shallow-water theory: II. Mixed Barotropic-Baroclinic Instability}}

\author{O. M. Umurhan\inst{1-3}}

   \offprints{O.M. Umurhan \email{oumurhan@ucmerced.edu}}

   \institute{
   School of Mathematical Sciences, Queen Mary
   University of London, London E1 4NS, U.K.\
     \and
        School of Natural Sciences, University of California Merced,
      Merced, CA 95343, USA\
           \and
        Astronomy Department, City College of San Francisco,
      San Francisco, CA 94112, USA\
}

\date{}
%{Received ------------- ; accepted -------------}

% 5 {} token are mandatory
%Calculations for laboratory
% experiments show that saturation of the instability is achieved through
% modification of the background shear when P$_{{\rm m}}$ is very low
 \abstract
     {Potential vorticity dynamics for astrophysical disks. }
      {Extend exploration of these instabilities in cold astrophysical disks which are  three dimensional whose mean states
      are baroclinic.  In particular, we seek to demonstrate the potential
      existence
      of traditional baroclinic instabilities of meteorological studies.  
      We show this in a simplified two-layer Philips Disk Model of a midplane symmetric disk.}
  {We consider the dynamical normal-mode response of thin annular
  disks with two potential
  vorticity defects, one in each of two disk layers. Each disk
  layer is of constant but differing densities.
  The resulting
  mean azimuthal velocity profile shows a variation in the vertical 
  direction implying that the system is baroclinic in the mean state. 
  The stability of the system is treated in the context of disk shallow
  water theory wherein azimuthal disturbances are much longer than the
  corresponding radial or vertical scales.  The normal-mode problem is solved
  numerically using two different methods.}
     {The results of a symmetric single layer barotropic model is considered
     and it is found that instability persists for models in which the potential vorticity
     profiles are not symmetric, consistent with previous results.  The instaiblity
     is interpreted in terms of interacting Rossby waves.  For a two layer
     system in which the flow is fundamentally baroclinic we report here
     that instability takes
     on the form of mixed barotropic-baroclinic type: instability occurs but
     it qualitatively follows the pattern of instability found in the barotropic models.  Instability
     arises because of the phase locking and interaction of the Rossby waves between
     the two layers.  The strength of the instability weakens as the density contrast
     between layers increases.}
     {Baroclinic instability is feasible for astrophysical disks but
     has the character of mixed barotropic-baroclinic type.
     The instability as
     explored in this study could be present in protoplanetary disks
     with weak vertical density stratification.  }

  %   {Nonsense}
  %   {Nonsense}
  %   {Nonsense}
  %   { Nonsense}
  %  {Nonsense}

\titlerunning{Barotropic-Baroclinic Instability}


  \maketitle
%________________________________________________________________________

\section{Introduction}
The phenomenon of baroclinic instability has been studied with intensity
for over 70 years.  Sometimes referred to as sloping 
convection (Vallis 2006), it is an instability that relies
on the presence of a vertical shear in a zonal flow
in a rotating atmosphere.  This archetypical sheared flow exists
as a result of a thermal-wind relationship balancing Coriolis effects
with meridional (equator-to-pole) temperature gradients.
It is ultimately a reflection of the mismatch in the isobars
and isopycnals in an atmosphere.
It is alternatively interpreted
as an instability resulting from the interaction of inter-layer
Rossby waves (Hoskins et al. 1985, Baines and Mitsudera 1994, Heifetz 2004).
Baroclinic instability is important to a variety of geophysical and astrophysical
phenonomena.  
Among others, it plays a well-known role in the generation of terrestrial weather
as well as in
the dynamics of the Ocean Mixed Layer (Haine and Marshall 1998).
It has been conjectured to play an important role 
in the mixing processes in stars (Knobloch and Spruit 1982, Fujimoto 1988).\par
The original studies of the instability (e.g. Charney 1947,
Eady 1949, Philips 1954) considered the process in the small
Rossby number limit in which the Rossby number, Ro, is given by
${\cal U}/2\Omega_0 {\cal L}$, where the terms are (respectively)
the typical speed of a vortex, planetary rotation rate (at latitude)
and the typical horizontal length scale of the flow.  
Baroclinic instability is a process that figures prominently
in the geostrophic limit (Ro $\ll 1$) but theoretical
considerations have shown that it persists relatively unabated
even when Ro is $ \sim 1$  (Stone 1970, Stone 1971, Molemaker 
et al. 2005).  These studies also argue that the instability
also persists for significant deviations from hydrostasy.
\par
The possibility that baroclinic instability is present in
accretion disk flows, which are characterized by Ro $=0.75$, 
was examined in the studies of Cabot (1984)
and  Knobloch and Spruit (1985,1986).
Owing to the fundamental inseparability of the resulting linearized
problem (even in the relatively simplified Boussinesq limit)
Knobloch and Spruit (1986) were only able 
to conclude that the instability is feasible
for radial and vertical disturbances comparable to the
local scale height of the disk.  The feature
complicating the analysis and rendering it
inseparable is the dominance of the Keplerian shear
making traditionally employed small Ro analyses of meteorological
studies here unusable. 
\par
Adding to the challenge: because the Keplerian shear is barotropic
the resulting problem for a disk is fundamentally
of mixed baroclinic-barotropic type.  This type of
mixed problem has been examined in the context of 
quasigeostrophic flows by 
McIntyre (1970), Ioannou \&
Lindzen (1986), James (1987).  These studies
generally agree that the structure of baroclinic waves
will get increasingly confined in the shearwise direction
with increasing strength of the barotropic shear.
The confinement, due
to the barotropic governor mechanism
(James 1987), appears to have the effect of
weakening growth rates.\par
Keplerian disks are notoriously difficult to finesse
into a geostrophic-hydrostatic framework familiar in traditional meteorology.  
 Knobloch and Spruit (1985,1986) note that it is only
in the long stremwise wavelength limit (azimuthal in disk geometry)
of a Keplerian flow will some of the formalism of quasigeostrophy
carry over.  Umurhan (2008) developed a scaling analysis in which
a local annular disk section can be viewed in a semi-geostrophic,
quasi-hydrostatic framework (see also Balmforth and Spiegel 1996). 
This, in turn,
permits a dynamical analysis purely in terms of the potential vorticity,
typical of meteorological studies.  This construction was used
to re-examine
the Rossby wave instability (Li et al. 2000, Meheut et al. 2010) 
by illustrating it in a setting stripped down to its essential components
(Umurhan 2010).
\par
We use the limit developed in Umurhan (2008) to examine
baroclinic instability in an annular section of a cold
accretion disk.  We circumvent the difficulties encountered
in Knobloch and Spruit's calculation by constructing
a two-layer Philips model (Philips 1954).  The Philips
model is an effective simplification of the 
governing equations that allows for analytical tractability
without losing the underlying physics at play
in the baroclinic instability (see an extended discussion of this
in the text of Vallis 2006).\par
This work is organized as follows.  In Section 2 we develop 
a two-layer Philips disk model from the disk shallow water
equations found in Umurhan (2008) utilyzing a number of
simplifications and assumptions.  In Section 3 we
set up the general normal mode problem for this
two-layer, mid-plane symmetric, model system including the boundary conditions
and numerical methods employed.  We describe the model
whose stability we test including a description
of the features we include which is designed
to strip away the complications arising from
critical layers.
In Section 4 we
consider the stability properties of a purely barotropic (single-layer)
configuration and interpret the results from the 
perspective of interacting Rossby waves developed in
previous studies. In Section 5 we build upon
the picture developed for the barotropic calculation
and describe a baroclinic
model configuration and examine its stability properties.
  In Section 6 we offer a summary of our principle results
  and a short litany of reflections and remarks.

\section{Derivation of the Two-Layer Philips Disk Model}

\begin{figure}
%\begin{center}
\leavevmode
\includegraphics[width=9.cm]{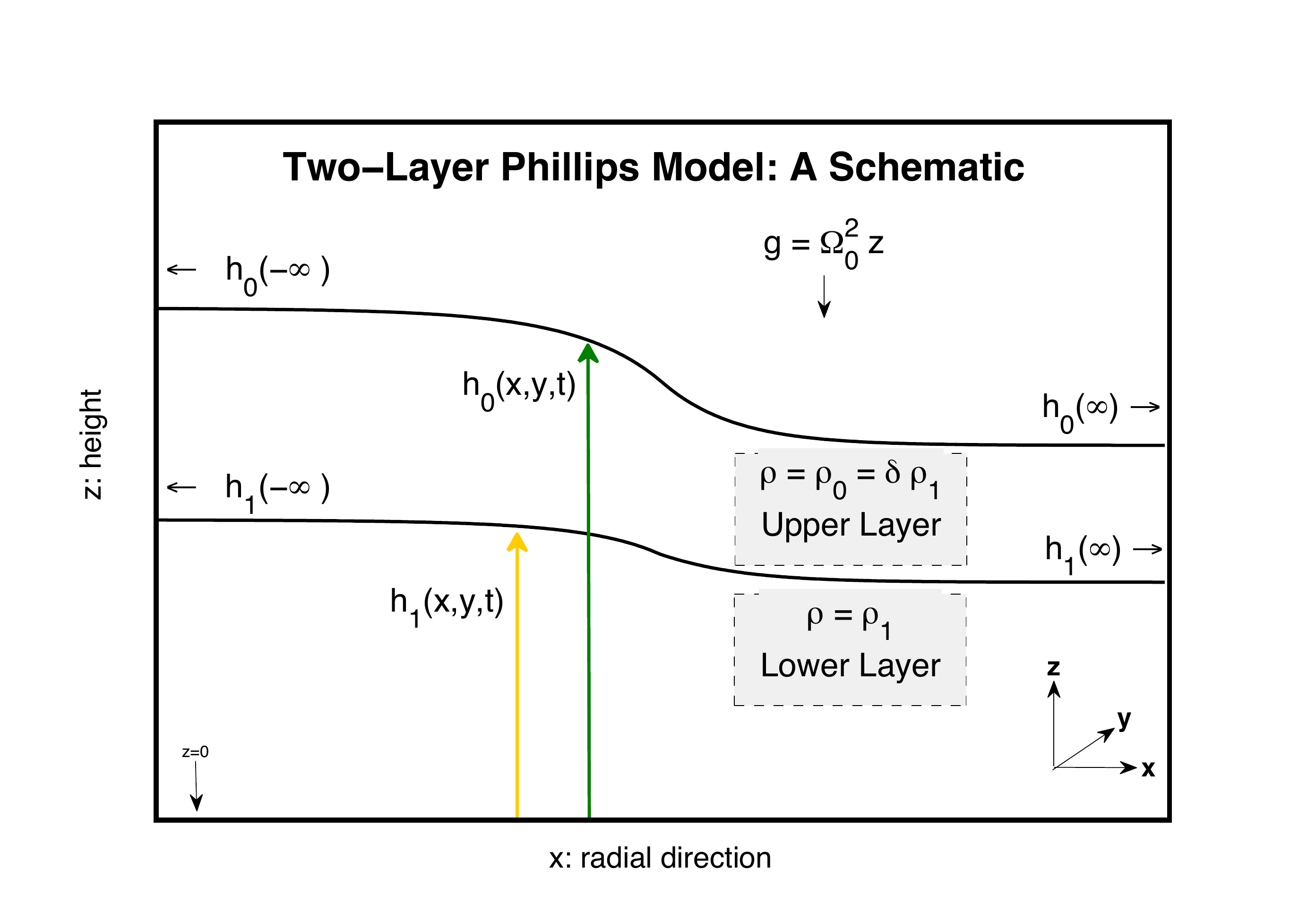}
%\end{center}
\caption{Configuration of two layer Philips Model.}
\label{Figure1}
\end{figure}

The analysis of this study assumes dynamics occur on azimuthal length scales
which are much longer than the system's vertical and radial scales.  The equations
appropriate for such a thin annular setting were developed from
a scaling analysis detailed in Umurhan (2008).
However they can be found in various other forms in older 
studies as well, e.g. Knobloch and Spruit (1985,1986)
and Balmforth and Spiegel (1996). \par
These equations describe dynamics in a thin annular section of
a disk centered
at some radius $R$ rotating with the local Keplerian rate $\Omega_K
= \sqrt{GM/R^3}$.  If the local soundspeed of the disk is given
by $c_s$ then the ratio $\varepsilon = c_s/(R \Omega_K)$ forms
a natural small parameter of the disk equations.  Formally
this means we will take $\varepsilon \ll 1$.  The equations
are therefore derived by assuming a number of scalings, namely
that the azimuthal velocity scales as $c_s$ and that the
vertical and radial velocities scale as $\varepsilon c_s$.  Together
with the assumption that the radial and vertical scales 
are a factor of $\varepsilon$ smaller than the azimuthal scale $R$,
the velocity assumptions here imply a long time
scale ${\cal T} \sim 1/\varepsilon\Omega_K$.  It is important
to note that this long timescale necessarily filters out
acoustic and gravity wave type motions from these dynamics.
The combination of these
scalings into the fundamental equations of motion
results in the following set of reduced non-dimensionalized equations
\beqa
-2\Omega_0 V &=& -\frac{1}{\rho}\partial_x P - q\Omega_0^2 x, \label{x_momentum} \\
\frac{d V}{dt} + 2\Omega_0 u &=& -\frac{1}{\rho}\partial_y P, \label{y_momentum}
\\
0 &=& -\frac{1}{\rho}\partial_x P + \Omega_0^2 z, \label{z_momentum}
\eeqa
together with
\beqa
\frac{1}{\rho}\frac{d \rho}{dt} + \nabla\cdot {{\bf u}} &=& 0, \label{continuity} \\
\rho T \frac{d S}{dt} &=& {\cal{Q}}, \label{heat_equation}
\eeqa
in which
\[
\frac{d}{dt} = u\partial_x + V\partial_y + w\partial_z.
\]
$P = \rho T$ is the pressure, where
$\rho$ and $T$ respectively denote density and temperature.  
The velocity vector ${\bf u} = (u,V,w)$
is comprised of components in the radial ($x$), azimuthal ($y$) and
disk vertical ($z$) directions respectively.  
{{The lengths $x$ and $z$
represent the radial and vertical scales
scaled by $\varepsilon R$ while the azimuthal
scales $y$ being scaled by
$R$.}}
The entropy
is $S \equiv \ln P/\rho^\gamma$ where $\gamma$ is the ratio of
specific heats.  $\Omega_0$ represents the non-dimensionalization
of the usual Coriolis terms that emerge when one moves into
a rotating frame.  In this formulation it is formally `$1$' but
we keep it around for the sake of latter clarity.  For Keplerian
disk systems $q= 3/2$.
\par
The lack of the usual
inertial terms in the resulting radial and vertical momentum
equations expresses the 
extremeness of the scalings describing
the thin annular disk region.
 The resulting equations describe dynamics
which are vertically hydrostatic but, moreover, geostrophic in the
radial direction.  In meteorological studies this is sometimes
referred to as semi-geostrophic because the azimuthal flow dynamics
are not in fundamental geostrophic balance.\par
To derive a Philips model that is tractable and transparent the
following assumptions are made:
\begin{enumerate}
\item Dynamics are symmetric about the midplane $z=0$,
\item Two layers are examined with differing uniform densities
(and the configuration is stable to buoyant instabilities),
\item The flow in each layer is incompressible. 
\end{enumerate}
The assumption of incompressibility means that
the continuity equation is replaced, instead, by\beq
\nabla\cdot{\bf u} = 0,
\eeq
in each layer.
A pictoral representation of the system envisioned
for this analysis is shown in Figure \ref{Figure1}.
From here on forth, quantities in the upper layer are designated
with a subscript `$0$' while in the lower layer they are
represented by the subscript `$1$'.  Thus, for example, the uniform
density of the upper layer is $\rho_0$ while in the lower layer
it is $\rho_1$.  Because incompressibility has been assumed
for each layer there is no longer any need for the energy
equation (\ref{heat_equation}).  Instead, the evolution of each layer's
thickness
will be followed (see below).\par
A qualitative inspection of the state of affairs depicted in 
Figure \ref{Figure1} shows that, indeed, the isopycnals and isobars
of the atmosphere are misaligned which means that the basic state
will be baroclinic with the in-layer azimuthal velocities being
different from one another.  The misalignment has been placed
here by hand and is controlled by the functional form of the heights $h_0$
and $h_1$.  The origins of their respective distributions 
ultimately rests upon
radially dependent thermal process involving radiation 
and disk chemistry requiring detailed 
calculations (e.g. Turner et al. 2012)
without making the simplifying assumptions
we have made.  For the purpose of this model, how the atmosphere got
arranged into this configuration is not our concern: it is enough that 
the profile is in steady state.\par

The azimuthal velocity in both layers are written as a sum of the local
Keplerian part $-q\Omega_0 x$ plus a deviation $v_i$ where
$i$ is either `$0$' or `$1$' as in,
$V_i = -q\Omega_0 x + v_i$.  
Assuming that the pressure of the disk is zero
at its top, the vertical momentum equation may be
immediately integrated to yield,
\beq
P = \Omega_0^2\frac{1}{2}\left\{
\begin{array}{ll}
\rho_0(h_0^2-z^2); & h_1\le z \le h_0, \\
 & \\
\rho_0(h_0^2-h_1^2) + 
\rho_1(h_1^2-z^2); & 0\le z \le h_1. 
\end{array}
\right.
\eeq
Putting this solution for the pressure back into
the horizontal momentum equations shows that
for each layer
\beqa
-2\Omega_0 v_i &=& -\partial_x \Pi_i, \label{v_eqns} \\
\left(\frac{D}{Dt}\right)_i v_i + \Omega_0(2-q) u_i
&=& -\partial_y\Pi_i, \label{v_evolution}
\eeqa
where
\beq
\Pi_0 = \frac{1}{2}\Omega_0^2 h_0^2, \qquad
\Pi_1 = \frac{1}{2}\Omega_0^2 \Bigl [
\delta h_0^2 + (1-\delta)h_1^2 \Bigr ].
\label{Pi_eqns}
\eeq
The quantity $\Pi_i$ may be viewed as the analog of the streamfunction
in meteorological studies.
Each layer height, which is in effect
an integration constant, is designated by $h_i(x,y,t)$.
The parameter $\delta \equiv \rho_0/\rho_1$ measures
the ratio of the two densities. Henceforth,
since interest is in configurations which are buoyantly stable,
attention will be given only to those values of $\delta < 1$. 

\par
Inspection reveals that the resulting horizontal momentum equations
are independent of the vertical coordinate.  As such the time derivative
operator in each layer has been written as
\[
\left(\frac{D}{Dt}\right)_i
= \partial_t + u_i\partial_x + (v_i-q\Omega_0 x) \partial_y.
\]
The incompressiblity equations may be integrated in each layer
revealing
\beq
w = \left\{
\begin{array}{ll}
-z(\partial_x u_0 + \partial_y v_0) + \bar W_0; & h_1\le z \le h_0, \\
 & \\
-z(\partial_x u_1 + \partial_y v_1); & 0\le z \le h_1. 
\end{array}
\right.
\label{vertical_velocities_blah}
\eeq
The symmetry condition has been imposed upon the vertical velocity
profile by enforcing that its value is zero at $z=0$.
The velocity term $\bar W_0$ is determined by enforcing certain matching
conditions when following the motion of the individual interfaces. 
Since $\bar W_0$ plays no essential role in the subsequent linear theory
we do not detail its form here.  The motion of each layer edge
is viewed from both layers and is accordingly set equal to the vertical velocities
found in (\ref{vertical_velocities_blah}).  Combining this with the criterion that there is no layer 
separation reveals that
the dynamics of each layer's {\em thickness} is governed by
\beq
\left(\frac{D}{Dt}\right)_1 h_1 = -(\partial_x u_1 + \partial_y v_1) h_1,
\eeq
and
\beq
\left(\frac{D}{Dt}\right)_0 (h_0-h_1) =
-(\partial_x u_0 + \partial_y v_0) (h_0-h_1).
\eeq
The above equations may be combined and simplified following standard procedures
resulting in
\beq
\left(\frac{D Q_0}{Dt}\right)_0  = 0,
\qquad Q_0 \equiv \frac{\Omega_0(2-q) + \partial_x v_0}{h_0-h_1},
\label{Q0_eqn}
\eeq
in the upper layer while, for the lower layer,
\beq
\left(\frac{D Q_1}{Dt}\right)_1 = 0, \qquad
Q_1 \equiv \frac{\Omega_0(2-q) + \partial_x v_1}{h_1}.
\label{Q1_eqn}
\eeq
The above relate the fact that 
the potential vorticities in each layer, represented by
$Q_0$ and $Q_1$, are materially conserved by the flow.  
It should be noted here that even though the $Q_i$ are conserved
in each of their respective layers, because the same $Q_i$ involve quantities 
that develop in layers outside of their own means that
layers are coupled to one another.
\par
Equations 
(\ref{Q0_eqn}-\ref{Q1_eqn}) 
together with
(\ref{v_eqns}-\ref{Pi_eqns}) completely describe the model
system. It is referred to hereafter as the {\em two-layer Philips disk
model}.

\section{Steady states, their perturbations and numerical method}
As the purpose of this work is to demonstrate the possible normal-mode responses
for various configurations of this two-layer Philips model, in this section we develop
the formalism for the perturbative study
of generalized steady state configurations.  This steady states are
assumed to be radially dependent, however, perturbations of these
states are non-axisymmetric.  In the
subsequent two sections we will detail the normal mode response for a variety of
specific steady state profiles.
\par
For the sake of transparency in the following presentation
 $\Omega_0$ shall be set to its nominal value of $1$.  
 We study perturbations of already laid down potential vorticity
 profiles.  These steady profiles describe  
 mean azimuthal velocities which are deviations around the Keplerian
 base flow.  There are no mean radial velocities.
 This means in effect, that the mean PV profiles will depend only
 upon the radial coordinate $x$.  Therefore, assuming 
 $Q_0 = \bar Q_0(x)$ and $Q_1 = \bar Q_1(x)$ it follows
 from the relationships between $Q_i$'s and $\Pi_i(h_0,h_1)$
 found in (\ref{Q0_eqn}-\ref{Q1_eqn}) that
 \beqa
   {2(2-q) + \partial_x^2 
 \bar \Pi_0} 
 -2\bar Q_0 {(\bar h_0-\bar h_1)}
  &=& 0, \label{steady_Q0}
 \\
   2(2-q) + \partial_x^2 \bar \Pi_1
   -2\bar Q_1 \bar h_1
 &=& 0, \label{steady_Q1}
 \eeqa
 in which
 the relationships between $\Pi$ and $h_i$, found in (\ref{Pi_eqns}), 
 have been inverted:
 \beq
 \bar h_0 = (2\bar\Pi_0)^{1/2},\qquad
 \bar h_1 = 2^{1/2}\left(\frac{\bar\Pi_1 - \delta \bar\Pi_0}{1-\delta}\right)^{1/2};
 \eeq
 expressing the mean steady height's $\bar h_i$ in terms
 of the steady in-layer pressures $\bar \Pi_i$.
 In deriving
 the above expressions the following relationships have also been explicitly
 used,
 \beq
\bar v_0  = \frac{1}{2}\partial_x^2
 \bar \Pi_0, \qquad
 \bar v_1 = \frac{1}{2}\partial_x^2
 \bar \Pi_1.
 \label{steady_h_new}
 \eeq
 
 Linear perturbations of the Philips model system developed
 by writing for all dependent quantities `$f$'
  \beq
 F_i \rightarrow \bar F_i(x) + F_i'(x,y,t)
 \eeq
 where the primes denote perturbations.
 Equations (\ref{Q0_eqn}-\ref{Q1_eqn}) 
and
(\ref{v_eqns}-\ref{Pi_eqns}) are explicitly written out
\beqa
(\partial_t + \bar v_0 \partial_y - qx \partial_y)Q_0' + u_0'\partial_x \bar Q_0 &=& 0, 
\label{Q_0_prime_equation}
\\
(\partial_t + \bar v_1 \partial_y - qx \partial_y)Q_1' + u_1'\partial_x \bar Q_1 &=& 0,
\label{Q_1_prime_equation}
\eeqa
in which
\beqa
(\bar h_0-\bar h_1)Q_0' &=& \partial_x v_0' - \bar Q_0(h_0'-h_1'), \label{Q_0_prime}\\
\bar h_1 Q_1' &=& \partial_x v_1' - \bar Q_1 h_1', \label{Q_1_prime}
\eeqa
 together with
\beqa
2v_0' &=& \partial_x \Pi_0', \label{v_0_prime}\\
2v_1' &=& \partial_x \Pi_1', \label{v_1_prime}\\
h_0' &=& (2\bar \Pi_0)^{-\sfrac{1}{2}} \Pi_0', \label{h_0_prime}\\
h_1' &=& \frac{1}{2^{\sfrac{1}{2}}}\left(\frac{1}{1-\delta}\right)^{\sfrac{1}{2}}
(\bar\Pi_1 - \delta\Pi_0)^{-\sfrac{1}{2}}\Bigl(\Pi_1' - \delta\Pi_0'  \Bigr).
\label{h_1_prime}
\eeqa 
 The linear equations resulting from perturbations
 of (\ref{v_evolution}) are rewritten in order
 to express the perturbation radial velocity, $u_i'$, in terms
 of the other perturbation quantities,
 \beqa
 u_0' &=& -\frac{1}{(\bar h_0-\bar h_1)\bar Q_0}
 \biggl\{\partial_y \Pi_0' + \Bigl(\partial_t + \bar v_0 \partial_y
 - qx \partial_y\Bigr) v_0' \label{u_0_prime}
 \biggr\}, \\
  u_1' &=& -\frac{1}{\bar h_1\bar Q_1}
 \biggl\{\partial_y \Pi_1' + \Bigl(\partial_t + \bar v_1 \partial_y
 - qx \partial_y\Bigr) v_1' \label{u_1_prime}
 \biggr\}.
 \eeqa
 Inserting (\ref{Q_0_prime}-\ref{v_1_prime}) and (\ref{u_0_prime}-\ref{u_1_prime})
 into (\ref{Q_0_prime_equation}-\ref{Q_1_prime_equation}) reduces the system
 to two coupled PDE's for the variables $\Pi_i'$.  This is to say,
 \beqa
 & & \Bigl(\partial_t + \bar v_0 \partial_y - qx \partial_y\Bigr)
 \Biggl[\partial_x^2 \Pi_0' - \partial_x\Pi_0' \cdot D\ln \bar Q_0 \nonumber \\
& & \hskip 1.75cm - 2\bar Q_0 (h_0'-h_1')\Biggr]
 -2\partial_y \Pi_0'\cdot D\ln Q_0 = 0, \label{single_Pi_0_eqn}
 \eeqa
 and
  \beqa
 & & \Bigl(\partial_t + \bar v_1 \partial_y - qx \partial_y\Bigr)
 \Biggl[\partial_x^2 \Pi_1' - \partial_x\Pi_1' \cdot D\ln\bar Q_1 
 - 2\bar Q_1 h_1'\Biggr] \nonumber \\
& & \hskip 3.0cm -2\partial_y \Pi_1'\cdot D\ln\bar Q_1 = 0, \label{single_Pi_1_eqn}
 \eeqa
 where the equations relating the perturbation heights to the perturbation pressures
 are given in (\ref{h_0_prime}-\ref{h_1_prime}).  The symbol `$D$' denotes differentiation
 with respect to $x$.\par
 
 All perturbation quantities
 are assumed to be periodic in the azimuthal direction and, furthermore, we assume 
 normal mode temporal structure.  That is to say, the solution ansatz we adopt is
 \beq
 f' = \hat f(x) e^{ik(y-ct)} + {\rm c.c.},
 \label{normal_mode_ansatz}
 \eeq
 where $k>0$ is the azimuthal direction wavenumber and $c$ is the wavespeed.  The goal
 will be to find the admitted values of $c$ for the variety of configurations examined 
 in the following sections.  Values in which Im$(c)>0$ correspond to instability.\par 
 
 For the radial boundary conditions it will be assumed that the fluctuation perturbation pressures
 go to zero at symmetrically placed radial boundaries, i.e. that $\Pi_i' \rightarrow 0$ as $x\rightarrow \pm L$.  
 These boundaries $L$
 will be placed significantly far from the region where there are nontrivial variations
 in the mean flow.  Typical
 values taken are $L\sim 10$ (unless otherwise specified).  
 We are concerned with uncovering effects which are insensitive to the effects
 introduced by using such an artificial boundary condition like we have here.
 As such, in all instances we have verified that the growth rates we
 determine are in fact insensitive to the position of the boundaries themselves
 by recalculating the growth rates for
 several values of $L$ significantly greater than $10$.  Because all eigenfunctions
 we generate show exponential decay as one approaches these artificial 
 boundaries, we are confident that there does exist, in fact, a minimum value
 of $L$ beyond which the effects of the boundaries are unimportant in determining
 the dynamics we study.  (Except, however, see the special case discussed at
 the end of Section \ref{Baroclinic_Profiles}).
 \par
  
 The coupled linearized ordinary differential equations are solved using a two-step procedure.  
 The x-domain is discretized on a Chebyshev grid $x^{(n)}$.  The number of 
 grid points is $N$. The solutions are assumed to
 be represented by the corresponding Chebyshev decomposition on this grid:
 \begin{enumerate}
 \item {\it{Step 1.}}
 This step involves expressing the
 governing equations (\ref{single_Pi_0_eqn}-\ref{single_Pi_1_eqn}) and their
 operators in physical space.  For a solution to $\Pi_i'$
represented in physical space by $\Pi^{(n)}_{(0,1)}$ the two coupled equations
form a single matrix equation
\beq
-ic {\mathbf M} {\mathbf \Pi} = {\mathbf L} {\mathbf \Pi},
\label{numerical_system_eigenvalues}
\eeq
where  
\[ {\mathbf \Pi} = 
\Bigl[\Pi^{(1)}_0, \Pi^{(2)}_0,\cdots,\Pi^{(N-1)}_0,\Pi^{(N)}_0,\Pi^{(1)}_1,\cdots,\Pi^{(N)}_1
\Bigr]^{\mathtt{T}},
\]
in which the superscript $`\mathtt{T}'$ denotes the vector transpose operation.
${\mathbf \Pi}$ is a vector of length $2N$ while
${\mathbf M}$ and ${\mathbf L}$ are $2N\times2N$ matrices.  
${\mathbf M}$ and ${\mathbf L}$ are matrices containing the Chebyshev representation
of the linear operators of the problem.
The boundary
condition that $\Pi'_{(0,1)}$ is zero at the endpoints are implemented the following
way.  The rows corresponding to the boundaries
are replace by the equation
\[
\partial_t \Pi^{(1,N)}_{(0,1)} = -\varpi \Pi^{(1,N)}_{(0,1)},
\]
where we choose $\varpi$ to be a very large number (typically $\sim 10^{9}$).
We then determine the full eigenvalue set, $c^{(n)}$ using standard packages
for determining numerical eigenvalues as found in Matlab.  We note
that the four eigenvalues of $\order{\varpi}$ 
are discarded since they are unphysical.\footnote{The normal modes
associated with these eigenvalues physically represents a very fast
relaxation of the system to the boundary condition.  This scheme
is used widely in the fluid dynamics literature.}
The remaining eigenvalues and corresponding
eigenvectors represent the responses of the system.  The boundary
conditions are satisfied with an error of $\order{1/\varpi}$.
We assume
a value of $N=300$ which typically yields converged solutions.
\item {\it{Step 2}} We verify and further refine the solutions we determine
in Step 1 by solving (\ref{numerical_system_eigenvalues}) using
a standard Newton-Raphson-Kantorovich (NRK) method.  We input into the
procedure a solution vector ${\mathbf \Pi}$ together with its
corresponding eigenvalue $\omega$.   The error tolerance
required of NRK solution was that it is correct to
one part in $10^9$.
\end{enumerate}
All quoted converged solutions in
this study passed both of the outlined steps above.  Furthermore, we
checked resolution convergence using the NRK method by taking
a given solution determined on a grid of size $N$, interpolate it onto
a grid of size $2N$, and then using this resulting interpolated
solution as an initial guess for the NRK
method.  In rare instances we found solutions determined via Step 1
to have false unstable eigenvalues, i.e. Im$(c) \neq 0$. These were identified using
the grid refinement procedure: for those
spurious eigenvalues,  
Im$(c) \rightarrow 0$ asymptotically as $N\rightarrow\infty$.

   \begin{figure}
%\begin{center}
\leavevmode
\includegraphics[width=9.cm]{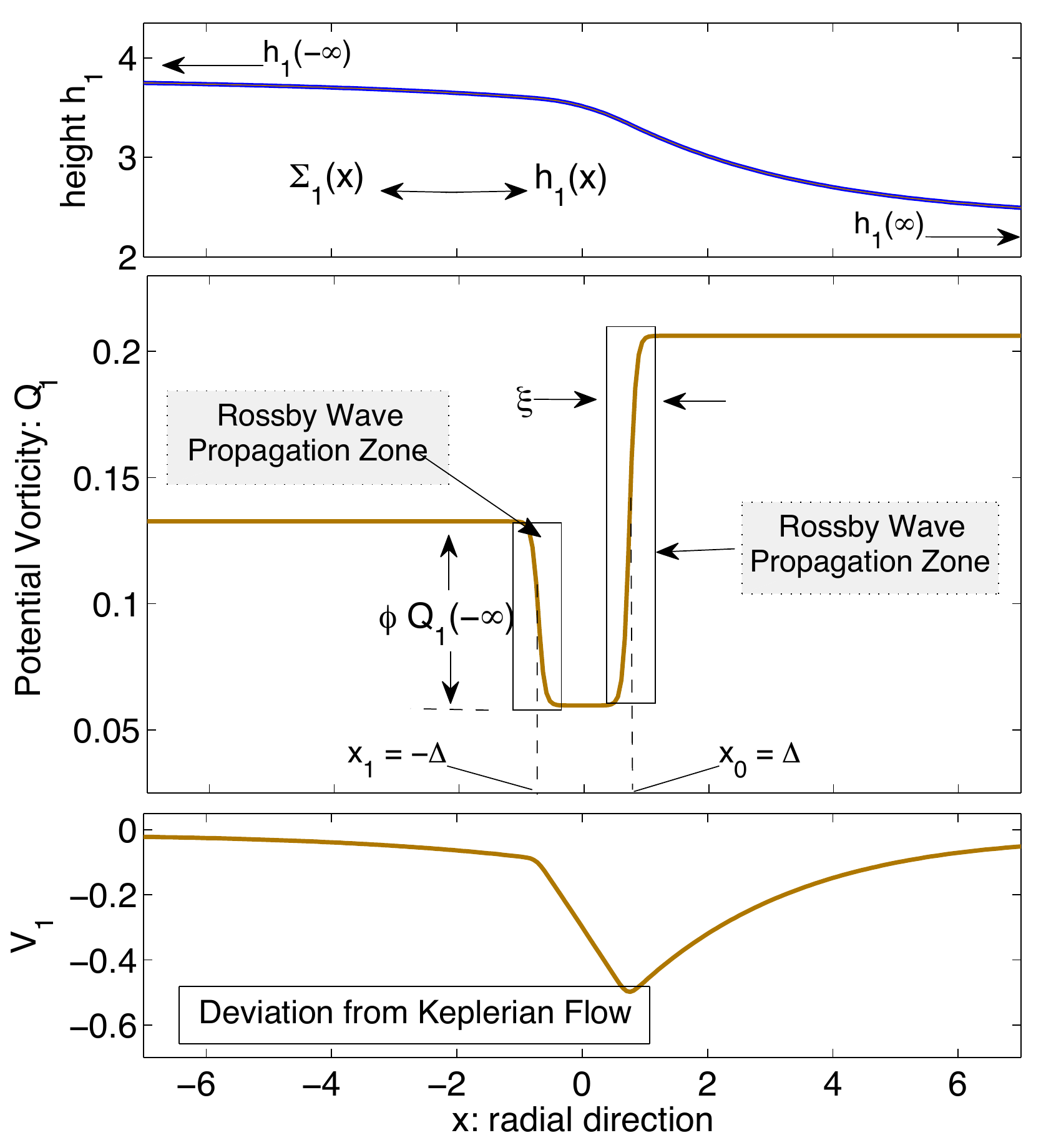}
%\end{center}
\caption{Single layer configuration and model.  Particular values
in reference to the `double-tanh' model (\ref{double_tanh_profile}): $h_1(\infty) = 2.4,
h_1(-\infty) = 3.77$ with $\Delta = 0.75$ and $\phi = \xi = 0.1, q = 1.5$.  The top
panel shows the corresponding mean height $h_1$.  The middle panel depicts the fundamental
PV profile together with descriptive annotations relating various features of the profile
to the parameters describing it in the text.  The bottom panel shows
deviation from Keplerian flow $\bar v_1$.
}
\label{single_layer_configuration}
\end{figure}
 
 \begin{figure}
%\begin{center}
\leavevmode
\includegraphics[width=9.cm]{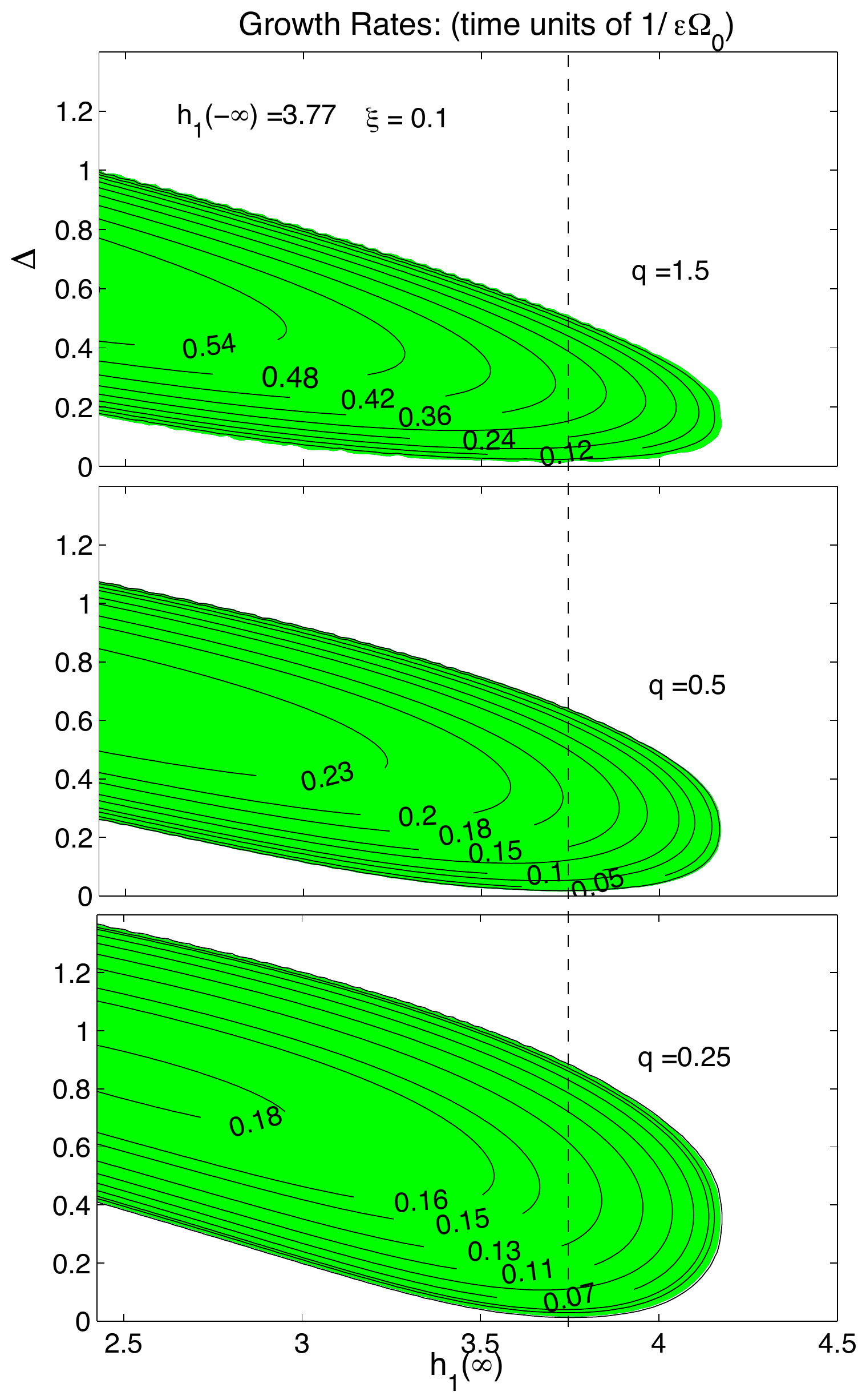}
%\end{center}
\caption{Single layer barotropic results for three different values of the background shear. 
Parameters of the `double-tanh' model (\ref{double_tanh_profile}): 
depth $\phi=0.1$, $h_1(-\infty) = 3.77, \xi = 0.1$.  
Growth rates shown as a function of varying the relative position
of the PV jump, $\Delta$, and variations of the asymptotic height as $x\rightarrow \infty$,
i.e. $h_1(\infty)$.  Instability growth rates in units of $1/(\varepsilon \Omega_0)$ (with
$\alpha = 1$). Instability
boundaries shown as shaded region. Top figure: $q=1.5$ (Keplerian). Middle and bottom figures $q=0.5,0.25$ 
respectively.  Vertical line designates symmetric step profiles.  As shear decreases
the corresponding growth rates decrease but the range of possible unstable values of $\Delta$
increases.}
\label{barotropic_results_1}
\end{figure}
 \section{Barotropic configurations}
 In order to anchor our intuition we begin this analysis by
 reconsidering the dynamics in a single-layer configuration.  Under
 those conditions the shear is barotropic.  
 The linearized equations we will need to solve are those given
 in the previous section with $\delta$ set to zero.    
 With $\delta = 0$ the upper layer has no dynamical effect upon
 the lower layer and, as such, the resulting dynamics describe
 flow of the lower layer entirely isolated from the upper layer.
 Therefore, the equations for the steady state
 are
 (\ref{steady_Q1}) with $\bar Q_1$ given and $\bar v_1$, as given in (\ref{steady_h_new}).
 For the linearized dynamics we solve only
  (\ref{single_Pi_1_eqn}) 
 with (\ref{h_1_prime})  to generate solutions for the
 lower layer perturbation enthalpy $\Pi_1'$.  
 We also assume the normal mode ansatz (\ref{normal_mode_ansatz})
 and enforce that $\Pi_1'(\pm L) = 0$.

 Figure (\ref{single_layer_configuration}) depicts
 the PV profile we shall examine here.
We refer to it as the `double-tanh' profile and is given
by the form
 \beqa
 \bar Q_1 &=& A_1\left[1+ \tanh\left(\frac{\Delta-x}{\xi}\right)\right] + \nonumber \\
 & & \hskip 2.0cm  B_1\left[1+ \tanh\left(\frac{x-\Delta}{\xi}\right)\right] + C_1,
 \label{double_tanh_profile}
 \eeqa
 where the constants are specified below.
 This functional form roughly represents a two-step
 PV-profile with the steps located at $x=\pm \Delta$ where $\Delta>0$
 is some parameter.  We will refer to twice this parameter as quantifying the 
 width of this type of PV profile. 
 {{The parameter $\xi$ represents the width of the transition in
 moving from one region of PV to the other.  In the limit
 $\xi\rightarrow 0$ the transition from one region of PV 
 to the other
 (e.g. near $x = \pm \Delta$ in Figure \ref{single_layer_configuration})
 is proportional to a Heaviside step function.  One one
 hand we numerically we try
 to avoid the $\xi\rightarrow 0$ limit, yet on the other hand
 we want to have $\xi$ be small in order to relate the results
 of the model to our simplified understanding of the system.
 As such we typically choose $\xi = 0.1$.
 Thus, as depicted in Figure \ref{single_layer_configuration},
 the limit $\xi\rightarrow 0$ makes the 
 PV-profile look like two successive
 step functions or `defects' in the PV.  We
 interpret the length $2\Delta$ as quantifying the 
 separation of these nominal defects of
 the PV.}}\par
 We must now relate
 the constants $A_1,B_1,C_1$ to physically intuitive features of the model.
 Because we envisage a PV profile that has an induced velocity profile (as
 a deviation from the background Keplerian flow, i.e. $\bar v_1$) that is zero
 as $x\rightarrow \pm \infty$, inspection of  (\ref{Q1_eqn})
 or (\ref{steady_Q1}) requires of us that
 \beq
 \lim_{x\rightarrow\pm\infty} \bar Q_1 = \frac{2-q}{h_1(\pm\infty)}.
 \eeq
  In other words, we design this profile 
 to yield constant mean heights, $h_1(\pm\infty)$, in the limit of
 $x\rightarrow\pm\infty$.  Of course, these heights may be different from
 one another but the constant asymptotic form of it ensures that
 the mean induced flow $\bar v_1$ correspondingly 
 asymptotes to zero as $x\rightarrow \pm \infty$.
 As such this implies that
 \beqa
 2A_1 + C_1 &=& \frac{2-q}{h_1(-\infty)}, \nonumber \\
  2B_1 + C_1 &=& \frac{2-q}{h_1(\infty)}. \nonumber
 \eeqa
 To complete this description we suppose that depth
 of the depression in $\bar Q_1$, located
 in the region $-\Delta<x<\Delta$,
 to be  some factor
 $\phi > 0$ of the asymptotic value of the PV as
 $x\rightarrow-\infty$, i.e.
 $\bar Q_1(-\infty)$.  
 This amounts to
 \[
 2A_1 - C_1 = \phi \bar Q_1(-\infty)= \phi\frac{2-q}{h_1(-\infty)}.
 \]
 \par
 In summary, then, the barotropic model (\ref{double_tanh_profile})
 is characterized by the five parameters: the asymptotic
 heights $h_1(\pm\infty)$, the symmetric positioning of the
 steps from $x=0$, $\Delta>0$, the relative depth of the PV-dip
 in the middle-zone of this model, $\phi$, and
 the width of the transition zones $\xi$.  Note that
 the middle-zone has a width equal to $2\Delta$.  The normal-mode
 system is additionally described by the shear parameter $q$.
  \par
  \bigskip
 A model similar to this one was considered in Umurhan (2010).
 Unlike here, however, the assumed PV-profile adopted
 in that previous study were (symmetric) step functions
 chosen to facilitate analytical tractability.  In comparison
 to the model considered here, the step function
 used in Umurhan (2010) would correspond to the
 limit of this model in which $\xi \rightarrow 0$. {Note also
 that the symmetry
 of the profiles studied in Umurhan (2010) is recovered here by
 choosing $h_1(\infty) = h_1(-\infty)$.}  
 We have therefore introduced
 the parameter $\xi$ to control the thickness of the 
 transition from one zone to the next.  \par
 The rationale
 of introducing this feature into the model is to 
 cleanly represent the propagation and interaction
 of Rossby waves without enduring the complexities
 of critical layers.  Because Rossby waves
 represent disturbances that propagate in regions
 where the PV changes, setting $\xi$ to be small
 ensures that the waves are {\em localized} to 
 those places where the PV changes the greatest.
 In this model, for example, there will be two
 Rossby waves, each of which are localized in $x$ 
 at $x=\pm\Delta$ with a radial width of scale $\xi$.  
  We typically use values of $\xi = 0.1$
 and this usually represented the Rossby wave dynamics
 without the appearance of critical layers.  The
 robustness some solutions was evaluated by also 
 generating solutions with values 
 of $\xi = 0.025$ (which necessitated using a higher
 resolution grid in order to resolve the transition zone
 with at least 5-7 grid points).
 Critical layers tended to appear for values of $\Delta \sim \xi$.
 However, we do not examine these situations in this study.\par
 The barotropic incarnation of the Rossby wave instability
 was interpreted in Umurhan (2010) to arise 
 as a result of the interaction of the two Rossby waves
 propagating along the azimuthal directions at the two locations, consistent with
 the more generic picture of interacting edgewave disturbances
 (Baines \& Mitsudera 1994).  An analogous form
 of this dynamic
 is present in this model and is especially
 pronounced for $\xi \ll 1$.
 For example, in reference to the PV profile shown in Figure
 \ref{single_layer_configuration}, we observe
 that the jump in the PV is positive at the step
 located at $x=1$. Therefore, in the local frame of reference
 of the flow at $x=1$ and {\em when  taken in isolation},
 the supported Rossby wave propagates
 in the negative $y$ direction.  
 Similarly, since the average change in the PV profile at $x=-1$
 is negative, it
 can support
 a Rossby wave propagating in the positive $y$ direction
 when viewed from the local reference frame of the flow
 at $x=-1$.  When the {\em laboratory measured} wavespeeds of
 both waves are nearly equal then instability may manifest
 itself, i.e. when the Hayashi-Young criterion
 is satisfied (Hayashi \& Young 1987).  Instability
 exists when the Rossby waves are
 essentially {\em counterpropagating} along the flow (Heifetz 1999) because
 it is only when the local propagating tendency of the Rossby
 wave is against the mean flow that the Hayashi-Young criteria of instability
 can be met.

 \par
 In the next section we will also refer to an effective
 surface density.  We define this here as
 \beq
 \Sigma_1(x) = \int_0^{h_1(x)} \rho_1 dz = \rho_1 h_1(x).
 \eeq
 As can be seen $\Sigma_1$ is equivalent to $h_1$ in the
 single layer model, but will not be so in the two layer model.
 Henceforth we take $\rho_1 = 1$.
\par
 A representative summary of the results of this section is displayed in Figure 
 \ref{barotropic_results_1}.  Even though we solve for
 the wavespeed $c$ we depict the growth rates Im$(-\omega) = {\rm Im}(k c)$,
 together with
 having here (and henceforth) set $k = 1$.\footnote{Disk-shallow water theory
 is such that the streamwise wavenumber drops out of all calculations
 owing to it being, essentially, a long-wavelenth theory.  See also
 Knobloch \& Spruit (1985,1986).}  
 The growth rates are quoted in time units used to derive the
 disk shallow water equations (see $\S 2$), namely, time is scaled in the longtime units
 $1/\epsilon\Omega_0$.
 \par
 The results we plot here are an extension
 of the symmetric models considered in Umurhan (2010).  Here we see that
 that the models do not have to display symmetric PV profiles in order for
 instability to surface and that, consistent with the results
 of Li et al. (2000), asymmetric PV profiles can just as easily 
 lead to instability.  We find that for a given set
 of layer parameters $h_1(\pm\infty)$ and $\phi$ (the depth of
 the PV depression) - as the background shear $q$
 is lessened, the range of separation values $\Delta$ that admits
 instability is shifted toward larger values.  In other words,
 weakening the shear means that profiles with steps further separated
 from one another will result in instability.  \par
 In terms of 
 the picture of interacting Rossby waves, this is easily
 rationalized in the way outlined earlier in this section: 
 (i) for instability to occur
 the laboratory measured phase speed of the Rossby waves
 along each step must be nearly the same, (ii) since
 the phase speeds of the Rossby waves are governed both by
 the background flow speed at the location of the step plus
 the intrinsic Rossby wave speed (as measured in the moving frame at the step) 
 induced by the local
 change in PV at the transition region it follows
 that, (iii)  with all else held equal, a reduction in the background shear 
 requires
 the location of the defects at $x=\pm\Delta$ 
 to be set further apart so that
 the laboratory measured phase speeds may be, once again, nearly equal. 
 \par
 When the step profiles are symmetric, i.e. when $h_1(-\infty) = h_1(\infty)$
 instability may occur for separation $\Delta = 0$ consistent with previous
 results in Umurhan 2010.  However, as $h_1(\infty)$ begins to deviate
 away from the value $h_1(-\infty)$ then there exists a minimum value
 for the defect separation above which instability may occur.
 It is also interesting to note that there is
 a maximum disparity of heights beyond which there is no instability
 possible. E.g. for values of $h_1(\infty)> 4.15$ with $q=1.5, \phi = 0.1$
 there is no instability possible in the model in Figure \ref{barotropic_results_1}.  The interpretation
 of this stability quality is once again similar to the above: varying $h_1(\infty)$
 beyond some critical value results in individual waves whose speeds
 cannot possibly match each other.  In particular, an increase of the speed 
 of the wave at the step $x=\Delta$ occurs
 because
 increasing $h_1(\infty)$ means that the average PV-gradient increases
 at $x=\Delta$ step which, in turn, 
 boosts the (in frame) counterpropagating Rossby wavespeed
 at the step.
 \par
 The insensitivity of these results on $L$ has been verified.  
 It is important to keep in mind that the growth rate values 
 convergence to the values quoted once $L$
 is greater than about 5.  For values of $L < 5$
 the growth rates weaken.  The presence of channel walls in problems
 like these creates the effect of 'image' waves (like image charges
 in electrostatics problems, Drazin \& Reid 1981) which act to inhibit phase locking.
 This behavior has been detailed in a study of the Rayleigh problem with 
 variable wall positions (Heifetz et al. 2009) and the trend for instability
 suppression with more narrowly set walls is consistent with their findings.
 
 The same character follows in
 the baroclinic cases studied except for one instance
 discussed later.
 Finally, consistent with the Howard Semicircle Theorem, we observe 
 that as the shear is weakened the maximum value of
 the growth rate is reduced.

\begin{figure}
%\begin{center}
\leavevmode
\includegraphics[width=8.cm]{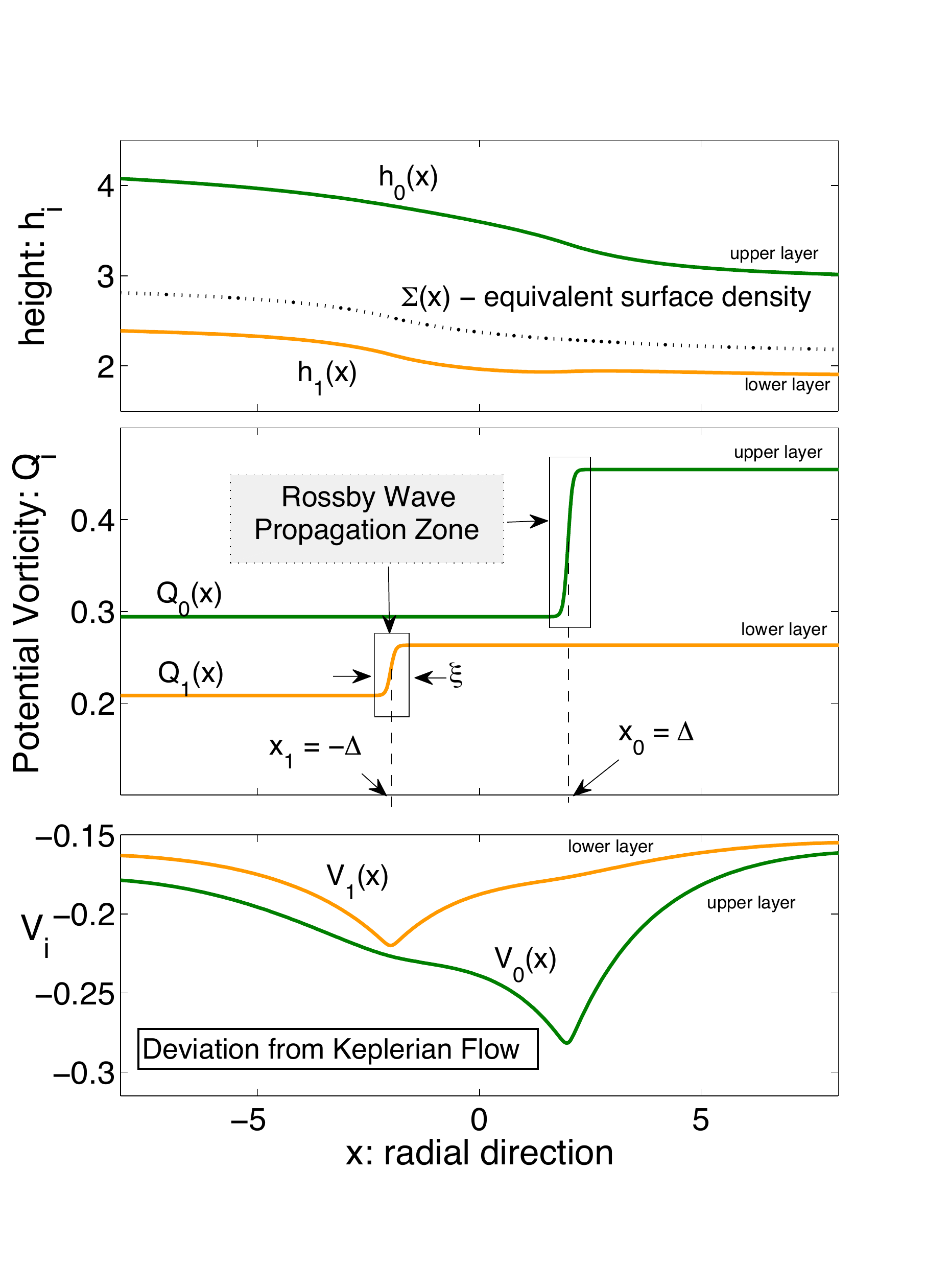}
%\end{center}
\caption{Baroclinic Jet profile model considered in this work.}
\label{jet_figure_1}
\end{figure}
\begin{figure}
%\begin{center}
\leavevmode
\includegraphics[width=9.15cm]{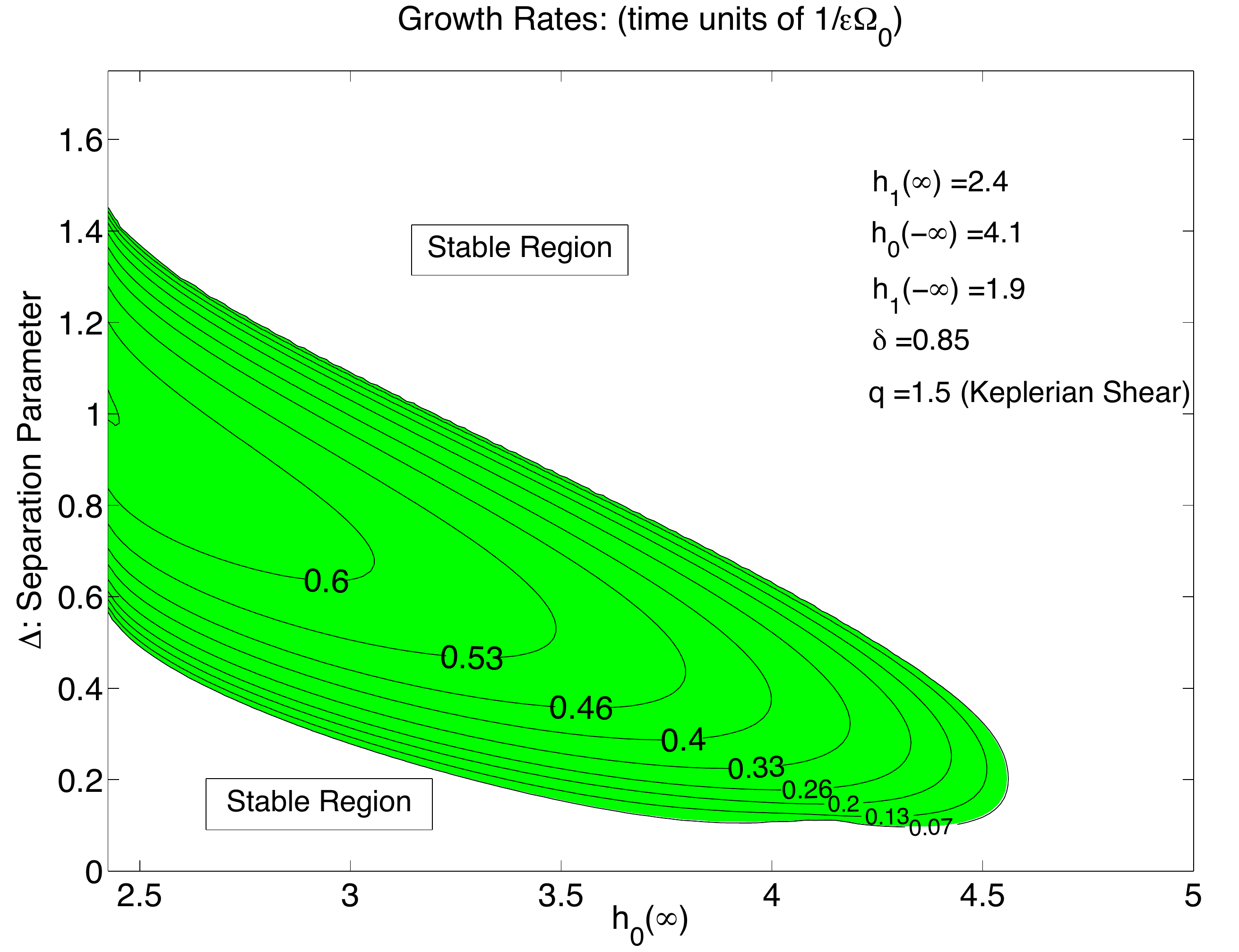}
%\end{center}
\caption{Mixed barotropic-baroclinic results.  The basic baroclinic
model is examined for variations in $\Delta$ and $h_0(\infty)$.  The
pattern of regions of growth is similar to similar barotropic configurations.
Note the particular absence of instability when $\Delta = 0$.}
\label{mixed_barotropic_baroclinic_1}
\end{figure}

\section{Baroclinic profiles}\label{Baroclinic_Profiles}
In this section we solve the two layer problem.  We assume that each
layer has a single jump in PV but that their positions in $x$ are, in general,
different from one another.  We employ the basic type of model encountered
in the previous section.  In particular we suppose that the mean PV
profiles
\beqa
\bar Q_0 &=& C_0 + A_0\left[1-\tanh\left(\frac{x-x_0}{\xi}\right)\right], \nonumber \\
\bar Q_1 &=& C_1 + A_1\left[1-\tanh\left(\frac{x-x_1}{\xi}\right)\right].
\eeqa
The steps for each layer are located at $x = x_0,x_1$ respectively.  Without loss of
generality we place these positions symmetrically about the point $x=0$ and assign
the position values $x_0 = -\Delta$ and $x_1 = \Delta$.  $\Delta$ is a parameter
as before, however, $\Delta$ can take on all values (and is not restricted to being strictly
positive).  As before, the thickness of the step in each layer is governed by
the parameter $\xi$.

\begin{figure}
%\begin{center}
\leavevmode
\includegraphics[width=9.15cm]{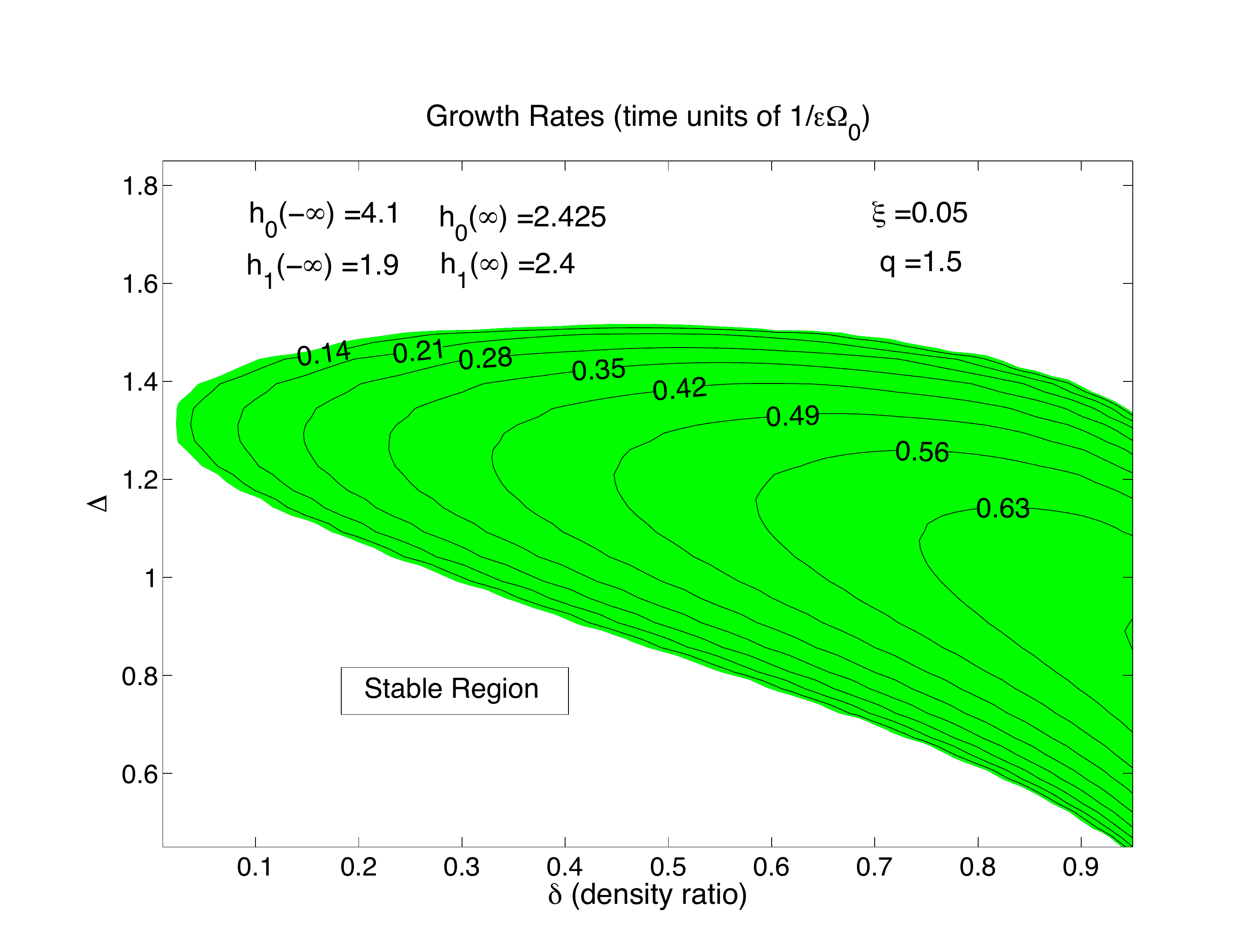}
%\end{center}
\caption{Baroclinic results depicting growth rates
for varying values of the density ratio $\delta$ against
the separation parameter $\Delta$.  Shown are the growth
rates for fixed values of $h_0(\pm \infty), h_1(\pm \infty)$
for Keplerian shear ($q=1.5$) and $\xi = 0.05$.  The range in 
separation values $\Delta$ in which instability occurs is
reduced as $\delta$ is reduced.  The growth rates
are reduced as well. The implications are that 
interlayer interaction is
weakened with a reduction of $\delta$.}
\label{mixed_barotropic_baroclinic_2}
\end{figure}

Following the arguments laid out in the previous section, we require
of the steady PV states to correspond to zero induced azimuthal
flow $\bar v_i$ as $x\rightarrow \pm \infty$.  Inspection of
(\ref{Q0_eqn}-\ref{Q1_eqn}) indicates that the constants $A_i, C_i$
must simultaneously satisfy
\beqa
& & C_0 + 2A_0 = \frac{2-q}{h_0(-\infty) - h_1(-\infty)}, \nonumber \\ 
& & C_0 = \frac{2-q}{h_0(\infty) - h_1(\infty)}  \nonumber \\
& & C_1 + 2A_1 = \frac{2-q}{h_1(-\infty)}, \qquad
C_1 = \frac{2-q}{h_1(\infty)}, \nonumber
\eeqa
where the parameters $h_i(\pm\infty)$ are the asymptotic values of the layer
heights in the appropriate limit of $x$.  The only requirement we place
upon the these are that $h_0(\infty)>h_1(\infty)$ and
$h_0(-\infty)>h_1(-\infty)$.\par
We solve for the steady configuration (\ref{steady_Q0}-\ref{steady_h_new})
using the above model for the PV profiles, which are governed by the six parameters
$h_i(\pm\infty),\Delta,\xi$.  The parameters of the steady
configuration are also determined by the three remaining other
parameters: the shear $q$, the density ratio $\delta$ and the domain size $L$.
A qualitative example of the resulting states
is shown in Figure \ref{jet_figure_1}.  It is evident from the
resulting flow that the mean deviation velocity fields are different
from one another between layers indicating that the resulting
flow state is baroclinic.  
\par
Figure \ref{mixed_barotropic_baroclinic_1} depicts the character
typifying instability in these models.  The parameters chosen
here were meant to motivate maximum comparison to the results
shown in the top graph of Figure \ref{barotropic_results_1}. 
The top layer has a density ratio $\delta = 0.85$ of the bottom
layer.  The results are qualitatively similar to the behavior
of the barotropic model with certain modification to the
region and degree of instability.  Since, by design, there are
only two waves which can interact - one emanating from the step
in the upper layer and the other from the step in the lower layer -
the interpretation of the instability is the same as for the
barotropic case.  Thus, baroclinic instability in this kind of
baroclinic problem
closely shadows the character of the instability in the barotropic
case.  The fundamental difference is, of course, the fact
that the source of the Rossby waves are from different vertical layers.
\par
The coupling between layers is governed by the density contrast $\delta$.
In Figure \ref{mixed_barotropic_baroclinic_2} we display
how the growth rates are modified for a model in which we
vary the separation $\Delta$ and the density contrast $\delta$.  Indeed, as one
might expect,  both the occurrence
of instability and its corresponding growth
rate diminish as $\delta \rightarrow 0$.  When viewed from
the perspective of a normal mode problem, the lower layer
essentially experiences no dynamical influence of the upper
layer when the mass of the upper layer is vanishingly small
(the lower layer ``knows" of the upper layer
due to the pressure fluctuations the upper layer induces
upon the lower layer).
From the mathematical viewpoint
of a normal mode problem, both layers decouple from
one another
when $\delta = 0$ (see the previous section).  Nonetheless
even for $\delta \approx 0$
the Rossby waves in each layer are finite (and non-zero) 
and propagate in isolation.
Thus when $\delta$ is increased the bottom layer feels
the upper layer in proportion to $\delta$ while the
upper layer feels the bottom layer by the latter's influence
upon the upper layer's thickness.  
\footnote{We note in passing that when treated as an {\it initial
value problem} the situation is different for $\delta$ nearly zero.
The lower layer still dynamically evolves free of the influence
of the upper
layer but the upper layer's dynamics {\em are} influenced by the
motion of the lower layer as it shows up in the upper layer's
dynamical evolution as an external forcing term.  We do not
take up this issue in this study.}
We also note here that the figures
show that growth rates persist
as $\delta \rightarrow 1$ but we are less inclined to
infer very much about this limit because 
as as $\delta \rightarrow 1$
the fundamental
scaling relations leading to the Philips disk model
begins to cause a break in the scaling assumptions
behind the disk shallow water equations.
Nevertheless the propensity for instability in these
baroclinic models for $\delta$ not very close to zero
suggests that baroclinic instability
may be manifest in disks which are weakly stratified.
We elaborate more upon this in Section 6.
\par
\begin{figure}
%\begin{center}
\leavevmode
\includegraphics[width=8.9cm]{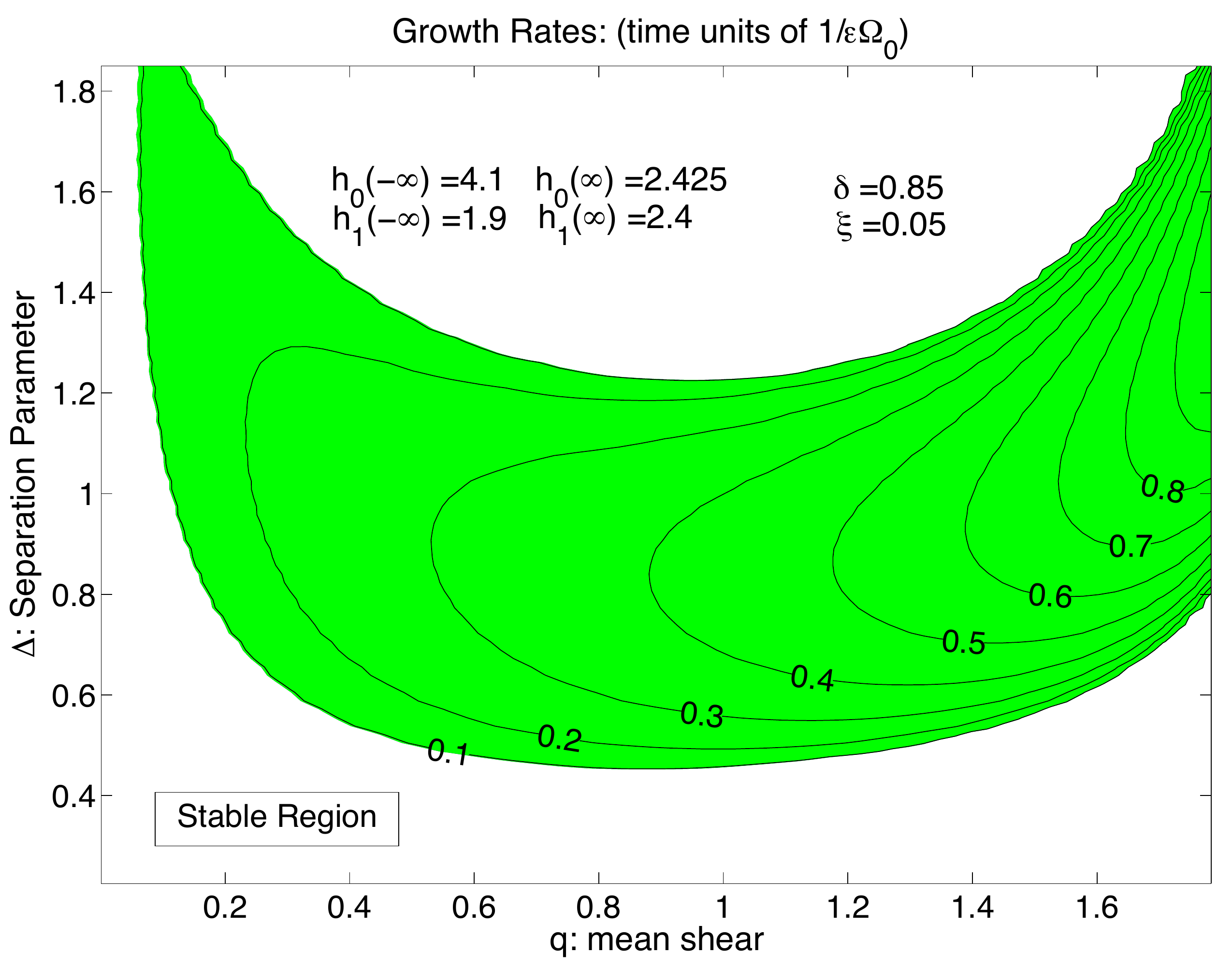}
%\end{center}
\caption{Baroclinic results depicting growth rates
for varying values of the shear parameter $q$ against
the separation parameter $\Delta$ with similar parameters found in
the previous figure.  As the shear tends to zero the values
of $\Delta$ for instability increases.  The growth rates
also tend to zero.
Growth rates become less reliable in the Rayleigh limit,
$q\rightarrow 2$ (see text).}
\label{mixed_barotropic_baroclinic_3}
\end{figure}

For a given set of model parameter values, varying
the shear $q \rightarrow 0$ acts to 
reduce the growth rate of the instability and
to push the unstable range of step separations $\Delta$
toward larger values.  In Figure \ref{mixed_barotropic_baroclinic_3}
we depict the growth rate trends for for variable
values of $q$ and $\Delta$.  Although the pattern of
instability in this parameter diagram is more complex
than the previous graphs, we see clearly that as
the shear weakens (holding the separation fixed)
the instability generally goes away.  On the other
end of the spectrum, as the Rayleigh limit is approached, i.e.
$q\rightarrow 2$, the instability also appears to become
suppressed.  However we note here that the scaling ordering
that is used to develop the disk shallow water equations
begins to breakdown as $q$ approaches $2$ (see Umurhan 2008).  
We therefore consider these observed trends (in this particular limit)
with more caution.  A proper investigation of what happens
in the Rayleigh limit should involve re-examining the equations
of motion in that particular limit. 
\par
\begin{figure}
%\begin{center}
\leavevmode
\includegraphics[width=9.15cm]{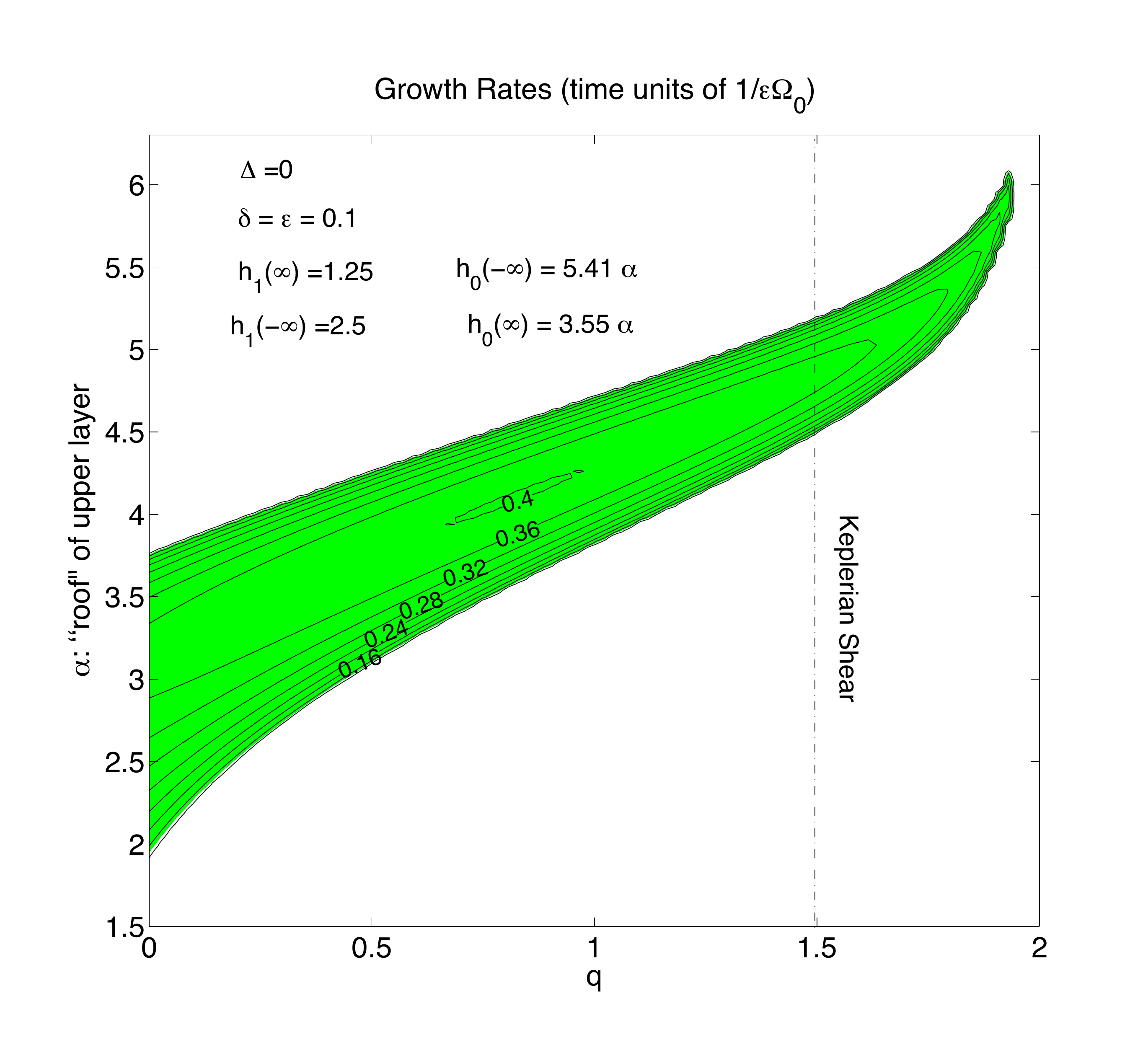}
%\end{center}
\caption{A model admitting baroclinic instability in which
the jets in each layer lie exactly on top of each other ($\Delta = 0$).
The plot depicts the variation of the growth rates as a function
of the shear $q$ and the so-called ``lid" of the atmosphere $\alpha$
(see text).  For this atmosphere model the radial
boundaries are set at $L=10$.  Keeping all parameters fixed but
rerunning this atmosphere model with $L=15$ reveals that all of
the instabilities depicted in this figure vanish.
}
\label{tbc_growth_rate_2}
\end{figure}

For given values of the shear $q$ there
appears to be no parameter configuration that we could find
in which there is both instability with the jets lying exactly on top of each
other ($\Delta = 0$) and where the instability does not vanish
as $L\rightarrow \infty$.  Such a configuration might correspond
to a flow profile in which the jets model a baroclinically
unstable setup 
relevant to more ``traditional" baroclinic flow problems (like
for the Earth and
other Solar System planets).
As we noted earlier, all of the results
we report in this study are those in which the growth rates remain
unaltered as the artificial domain walls in $x$ are made very large
$L\gg 1$.  In Figure \ref{tbc_growth_rate_2} we show an example
set of parameters in which there is instability but where
the separation in $x$ of the two PV jumps is zero.  The values
of the top of the atmosphere $h_0(\pm \infty)$ are represented
in a way such that the ratio of $h_0(\infty)/h_0(-\infty)$ remains
fixed but the overall height of the atmosphere can vary with the
parameter $\alpha$.  For the example depicted, $h_0(\infty) = 3.55\alpha$
and $h_0(-\infty) = 5.44\alpha$.  Thus $\alpha$ acts like a parameters
varying the lid of the atmosphere.  The values of the lower
layer parameters $h_1(\pm\infty)$ are fixed.  As such
Figure \ref{tbc_growth_rate_2} displays the growth rates as a function
of $\alpha$ and $q$ with $L=10$.  Instability appears to exist
but for values of $\alpha$ so large that the atmosphere
is effectively thin and tall.  When
we then increase the value of $L$ (holding fixed the remaining
atmosphere parameters) the instability eventually vanishes.
\par
We therefore conclude that the possibility of instability
in a model in which the jets lie on top of each other,
e.g. as implied by the results shown in Figure \ref{tbc_growth_rate_2},
is an artifact of the narrowness of the
radial boundaries and is not likely to be possible for
a real protoplanetary disk.

\section{Summary and Discussion}
Baroclinic instability in the guise familiar in meteorology (in which 
isopycnals and isobars are misaligned) 
is feasible for
cold protoplanetary disks, albeit in a mixed baroclinic-barotropic form owing
to the strength of the underlying Keplerian shear.  By using a two-layer
Philips model we have further substantiated the conjecture made by
Cabot (1984) and Knobloch \& Spruit (1984,1985) that baroclinic instability
is relevant for accretion disks.  Aside from determining growth rates
and interpreting the mechanics of the instability in terms of
Rossby waves interacting across layers,
it is also shown that the instability weakens in strength as the vertical 
density
stratification of the layer is increased.   
The results of the two-layer model suggests that meteorological-type of
baroclinic processes, which wholly depend on the mismatch of
isobars and isopycnals in the mean state, is likely 
for disks that have weak vertical density stratification.  Of course, this
does not limit the full scope of possibilities and, as such, this 
must be analyzed more fully in the setting of more realistic disk models.

The use of the simple Philips two-layer model has been shown to be very good at
capturing the essence of baroclinic flow dynamics present in continuously layered
models of atmospheres (Vallis 2006).  Comparison of growth rates determined
for both the Eady Model (continuous vertical shear) and the Philips model shows that, 
although they are not precisely the same,
the qualitative behaviors are consistent with one another and it is for this reason
why the Philips model is so widely regarded as a useful theoretical platform to examine
the physics of these and related processes.  In fact, studies have been done in which multiple
layer Philips models have been examined for the same problem yielding no significantly
different results in classical meteorological investigations.  This has partly to do
with the geometry of the system being analyzed: in a planetary atmosphere there is usually
a bottom boundary and the direction of gravity is the same.  

By contrast we have investigated a disk model in which symmetry with respect to the disk
midplane is imposed a priori.  Thus the dynamics possible here, which allows only for 
a subset of possible phase relationships and Rossby wave interactions 
between the midplane layer and the layer
above it, is a subset of a wider class of dynamical responses possible.  For instance,
if the restriction of midplane symmetry were relaxed then there comes into existence
the possibility of new types of Rossby wave interactions independently involving three layers (the midplane and
the two layers sandwiching the midplane) which will contain the dynamics of 
the two-layer system studied here.  This may yield qualitatively new dynamical possibilities.
\par
{{We should note here the similarities between traditional
`meteorological'
baroclinic instabilities to other 
types of baroclinic processes investigated for both disks
 (e.g. Klahr \& Bodenheimer 2003, Petersen et al. 2007a
Petersen et al. 2007b, Lesur \& Papaloizou 2010,  Lyra \& Klahr 2011) and the Sun (Gilman \& Fox 1997,
Zaqarashvili et al. 2010, to name only a few studies).  In these analyses
 potential vorticity disturbances are effectively baroclinically
torqued by some process that is either thermal or magnetic
or, in some investigations, a combination of both. 
In those studies of magnetic Rossby waves in the Sun
(e.g. Gilman \& Fox 1997,
Zaqarashvili et al. 2010)  Rossby waves will propagate along those places where
there are strong gradients in the solar differential rotation.  
Treated as a hydrodynamic problem,
these Rossby waves 
are not necessarily unstable on their own, especially for differential
rotation profiles inferred to be characteristic of the Sun's tachochline.
(Spiegel \& Zahn 1992).
However,
when a strong toroidal field is added to the mix there appears
an effective baroclinic source term (owing to the Lorentz force)
which can render the wave unstable.  In all of these studies
the instabilities are global in character as 
is the instability investigated in this study 
since we have examined the fate of disturbances with 
azimuthal wavelengths on the scale of the disk.}}

\par
The upcoming Paper III in this series examines an alternative
derivation of the Shallow Water Equations for disks that lifts
the restrictions placed upon the azimuthal scales as
compared to the vertical and radial scales.


\begin{thebibliography}{}
%\bibitem[1972]{as72}
%Abramowitz, M \& Stegun, I. A., 1972, Handbook of Mathematical Functions,
%Dover, New York
\bibitem[1994]{baines94}
Baines, P. G., \& Mitsudera, H. 1994, J. Fluid Mech., 276, 327

\bibitem[1996]{balmforth96}
Balmforth, N. J., \& Spiegel, E. A. 1996, Phys. D, 97, 1

%\bibitem[1999]{bender_orszag_1999}
%        Bender, C. M., \& Orszag, O. 1999, Advanced Mathematical Methods for Scientists and Engineers %(Springer)

%\bibitem[1989]{bodo1989}
%        Bodo, G., Rosner, R., Ferrari, A. \& Kobloch, E. 1989,
%        ApJ, 341, 631

%\bibitem[1961]{chandra61}
%       Chandrasekhar, S. 1961, Hydrodynamic and Hydromagnetic Stability (Oxford)

\bibitem[1984]{cabot_82}
		Cabot, W. 1984, ApJ, 277, 806
		
\bibitem[1947]{charney_47}
Charney, J. G. 1947, J. Meteor.,
4, 135
		
\bibitem[1981]{drazin_81}
Drazin, P.G., \&  Reid, W.H. 1981 , Hydrodynamic Stability, Cambridge Univ. Press, Cambridge

%\bibitem[2005]{dubrulle}
%    Dubrulle, B., Marie, L. , Normand, Ch., et al. 2004, A\&A, 429, 1

\bibitem[1949]{eady}
Eady, E. T. 1949  Tellus, 1, 33

%\bibitem[2009]{ebrahimi09}
%        Ebrahimi, F., Prager, S.C., \& Schnack, D.D., 2009,
%        ApJ, 698, 233

%\bibitem[1990]{friedman56}
%        Friedman, B. 1956, Principles and Techniques of Applied Mathematics,
%        Wiley, New York

\bibitem[1988]{fujimoto_88}
		Fujimoto, M. Y. 1988
		A\&A, 198, 163
		
\bibitem[1997]{Gilman97}
		Gilman, P. A., \& Fox, P.A. 1997
		ApJ, 484, 439

\bibitem[1999]{godon}
        Godon, P., \& Livio, M. 1999, ApJ, 523, 350

%\bibitem[1986]{goldreich86}
%    Goldreich, P., Goodman, J., \& Narayan, R. 1986, MNRAS, 221, 339

%\bibitem[1965]{goldreich_lyndenbell_65}
%    Goldreich, P., \& Lynden-Bell, D. 1965, MNRAS, 130, 125
    
\bibitem[1998]{Haine_Marshall}
	Haine, T. W. N., \& Marshall, J. 1998, J. Phys. Ocean., 28, 634

    \bibitem[1985]{hayashi-young}
    Hayashi, Y.-Y., \& Young, W. R. 1987, J. Fluid Mech., 184, 477

\bibitem[1999]{Heifetz99}
Heifetz, E., Bishop, C. H., \& Alpert, P. 1999
Q. J. R. Meteorol. Soc., 125, 2835

\bibitem[2009]{Heifetz09}
Heifetz, E., Harnik, N. \& Tamarin, T. 2009
Q. J. R. Meteorol. Soc., 135, 2161

%\bibitem[1978]{LovelaceHohlfeld78}
%Lovelace, R.V.E., \& Hohlfeld, R.G. 1978
%ApJ, 221, 51

\bibitem[1985]{Hoskins85}
    Hoskins, B. J., McIntyre, M. E., \& Robertson, A. W. 1985
    Q. J. R. Meteorol. Soc., 111, 877
    
\bibitem[2003]{Klahr03}
    Klahr, H. H., \& Bodenheimer, P. 2003
    ApJ, 582, 869
    
\bibitem[1982]{knobloch_82}
	Knobloch, E., \& Spruit, H. C. 1982
	A\&A, 113, 261

\bibitem[1985]{knobloch_85}
	Knobloch, E., \& Spruit, H. C. 1986
	Geophys. Astrophys. Fluid Dyn., 32, 197
	
\bibitem[1986]{knobloch_86}
	Knobloch, E., \& Spruit, H. C. 1986
	A\&A, 166, 359

%\bibitem[1962]{Knauer66}
%    Knauer, W. 1966
%    J. App. Phys., 37, 602

\bibitem[2010]{Lesur}
	Lesur, G. \& Papaloizou, J. C. B. 2010
	A\& A, 513, 60
\bibitem[2000]{Li00}
    Li, H., Finn, J. M., Lovelace, R. V. E., \&  Colgate, S. A. 2000 (LFLC-2000)
    ApJ, 533, 1023

%\bibitem[2001]{Li01}
%    Li, H., Colgate, S.A., Wendroff, B., \& Liska R. 2001
%    ApJ, 551, 874

%\bibitem[2007]{lithwick07A}
%    Lithwick, Y. 2007
%    ApJ, 670, 789 (Lithwick 2007 A)

%\bibitem[2007]{lithwick07B}
%    Lithwick, Y. 2007
%     {\tt arXiv:0710.3868[astro:ph]} (Lithwick 2007 B)

%\bibitem{lovelace}
%    Lovelace, R. V. E., Li, H., Colgate, S. A. \& Nelson, A. F. 1999
%     ApJ, 513, 805

\bibitem{lyra}
	Lyra, W. \& Klahr, H. 2011
	A\&A, 527, 138
	
\bibitem{meheut}
    Meheut, H., Casse, F., Varniere, P., \& Tagger, M. 2010
    A\&A, 516, A31
    
\bibitem{molemaker_2005}
	Molemaker, M. J., McWilliams, J. C., \& Yavneh, I. 2005
	J. Phys. Ocean., 35, 1505

%\bibitem[2007]{muraki07}
%        Muraki, D. J. 2007
%        SIAM J. App. Math., 67, 1504

%\bibitem[1984]{PP84}
%        Papaloizou, J. C. B., \& Pringle, J. E. 1984
%        MNRAS, 208, 721

\bibitem[1987]{pedlosky}
        Pedlosky, J. 1987, Geophysical Fluid Dynamics (2nd ed.),
        Springer-Verlag, New York
        
\bibitem[2007]{petersenA}
Petersen, M. R., Keith J., \& Stewart, G. R. 
2007a, ApJ 658 1236

\bibitem[2007]{petersenA}
Petersen, M. R., Keith J., \& Stewart, G. R. 
2007b, ApJ 658 1252

\bibitem[1954]{philips_54}
Phillips, N. A. 1954, Tellus,6, 273

%\bibitem[1880]{Rayleigh1880}
%        Lord Rayleigh 1880
%        Proc. Royal Math. Soc., 9, 57


%\bibitem[1989]{salby}
%        Salby, M. L. 1989, Tellus, 41A, 48

\bibitem[1992]{spiegel}
		Spiegel, E. A., \& Zahn, J.-P. 1992, A\&A 265, 106
        
\bibitem[1970]{stone_70}
		Stone, P. H. 1970, J. Atm. Sci., 27, 721
		
\bibitem[1971]{stone_71}
		Stone, P. H. 1971, J. Fluid Mech., 48, 659

%        \bibitem[2007]{tevzadze07}
%    Tevzadze, A.G., Chagelishvili, G.D., Zahn, J.-P. 2008,
%    A\&A, 478, 9

%\bibitem[2004]{ur04}
%      Umurhan, O. M., \& Regev, O. 2004, A\&A, 427, 855

\bibitem[2012]{turner12}
 Turner, N. J.,  Choukroun, M.,   Castillo-Rogez J., \&  Bryden G.
2012,  Accepted ApJ
{\tt arXiv:1110.4166v3[astro:ph]}

\bibitem[2005]{as72}
Vallis, G. K. 2006, Atmospheric and Oceanic Fluid Dynamics,
Cambridge University Press, New York



\bibitem[2008]{umurhan08}
    Umurhan, O. M. 2008, A\&A, 489, 953
    
\bibitem[2010] {umurhan10}
      Umurhan, O. M. 2010, A\&A, ,521, A25
      
\bibitem[2010]{Zaqarashvili}
		Zaqarashvili, T. V., Carbonell, M., Oliver, R., \&
		Ballester, J. L. 2010
		ApJ, 709, 749
		

%\bibitem[2000]{yavneh00}
%Yavneh I., McWilliams J. C. \& Molemaker M. J. 2001 J. Fluid Mech., 448, 1


\end{thebibliography}
\end{document}